\documentclass[10pt,a4paper]{article}

\usepackage{times}

\usepackage{graphicx} 

\usepackage{fancyhdr}

\usepackage[lmargin=2.5cm, rmargin=2.5cm,tmargin=3.50cm,bmargin=3.50cm]{geometry}
\usepackage{indentfirst}

\usepackage{amsmath}
\usepackage{amssymb}

\usepackage{titlesec}
\titleformat{\section}[hang]
  {\centering}{\thesection}{1ex}{\normalsize \textsc}
\titleformat{\subsection}[hang]
  {}{\thesubsection}{1ex}{\normalsize \textit}

\renewcommand{\thesection}{ \normalsize \textnormal{\Roman{section}.}}
\renewcommand{\thesubsection}{\normalsize \textnormal{\textsc{\textit{\Alph{subsection}.}}}}

\def\e{\begin{equation}}
\def\f{\end{equation}}
\def\_#1{{\bf #1}}

\def\.{\cdot}

\begin{document}

\title{\large \textbf{Towards optimal spatiotemporal wavefront shaping for the cocktail party problem with inverse design of an acoustic reconfigurable metasurface in disordered media}}
%
\def\affil#1{\begin{itemize} \item[] #1 \end{itemize}}
\author{\normalsize \bfseries \underline{R. Pestourie}$^1$, C. Bourdeloux$^2$, F. Lemoult$^2$, M. Fink$^2$  and S. G. Johnson$^3$}

\date{}
\maketitle
\thispagestyle{fancy} 
\vspace{-6ex}
\affil{\begin{center}\normalsize $^1$School of Computational Science and Engineering, Georgia Tech, Atlanta, Georgia, USA\\
$^2$Institut Langevin, ESPCI Paris, Université PSL, Paris, France\\
$^3$Mathematics Department, MIT, Cambridge, Massachusetts, USA\\
rpestourie3@gatech.edu
 \end{center}}

\begin{abstract}
\noindent \normalsize
\textbf{\textit{Abstract} \ \ -- \ \
Multiple-user multiple-input multiple-output applications have recently gained a lot of attention. Here, we show an efficient optimization formulation for the design of all the temporal and spatial degrees of freedom of an acoustic reconfigurable metasurface for the cocktail party problem. In the frequency domain, the closed-form least square solution matches the optimal time reversal solution for multiple emitter-receiver pairs, optimizing for each frequency independently. This is more efficient than solving in the time domain where the time convolution mixes all the degrees of freedom into a resource-intensive optimization. We illustrate this methodology by optimizing the frequency response of a design for two pairs of emitters-receivers using the Green's functions of disordered media that are measured experimentally. We report strong performance that will be put in perspective in future work, where we will analyze the robustness of the design to noise in the data and design the convolutional filters that match the optimal frequency response for experiment validation of the design.}
\end{abstract}

\section{Introduction}
The cocktail party problem involves a complex wave scattering phenomenon that appears across many applications \cite{Bourdeloux24, Arons92}. Acoustic reconfigurable metasurfaces in reflection~\cite{Assouar18} can leverage disordered media to focus emitters’ signals on desired target receivers~\cite{Sheng06, Kaina14}. The intuitive design of time reversal has shown successes over broadband signals~\cite{Bourdeloux24}. Although time reversal is optimal in single input single output (SISO) settings, it is no longer optimal in multiple-user multiple-input multiple-output (MU-MIMO) settings~\cite{Montaldo04}. Past work has successfully used combinatorial greedy algorithms to optimize metasurfaces in MU-MIMO settings~\cite{Kaina14, Zhang24}, however, it may not scale to large metasurfaces and full temporal and spatial degrees of freedom. Here, we exploit the knowledge of the optimal time reversal solution to formulate the MU-MIMO design problem that we solve efficiently in the frequency domain. Given measurements of the Green’s functions in disordered media and using the fact that the convolution turns into a product in the frequency domain, we cast the MIMO design problem as a least square fit for each frequency independently and match the optimal time reversal solution for multiple emitter-receiver pairs. We solve for the optimal frequency response for each element of the metasurface in closed form, for large problems in space this step can be done in parallel for each frequency. We illustrate this methodology in MIMO settings using the experimentally measured Green's functions and report strong performance. However, the design's robustness to noise in the data needs to be further studied. Experimental validation in future work will require designing convolutional filters that generate these frequency responses using methods such as the Parks-McClellan algorithm~\cite{McClellan75}. However, note that the convolutional filter design of each metasurface element can be performed independently and in parallel. This efficiently solves the full optimization taking into account all space and time degrees of freedom compared to solving in the time domain, where the time convolution mixes all the time taps in a way that is hard to efficiently exploit computationally.

\begin{figure}[h!]
\centering \includegraphics[width=.85
\textwidth]{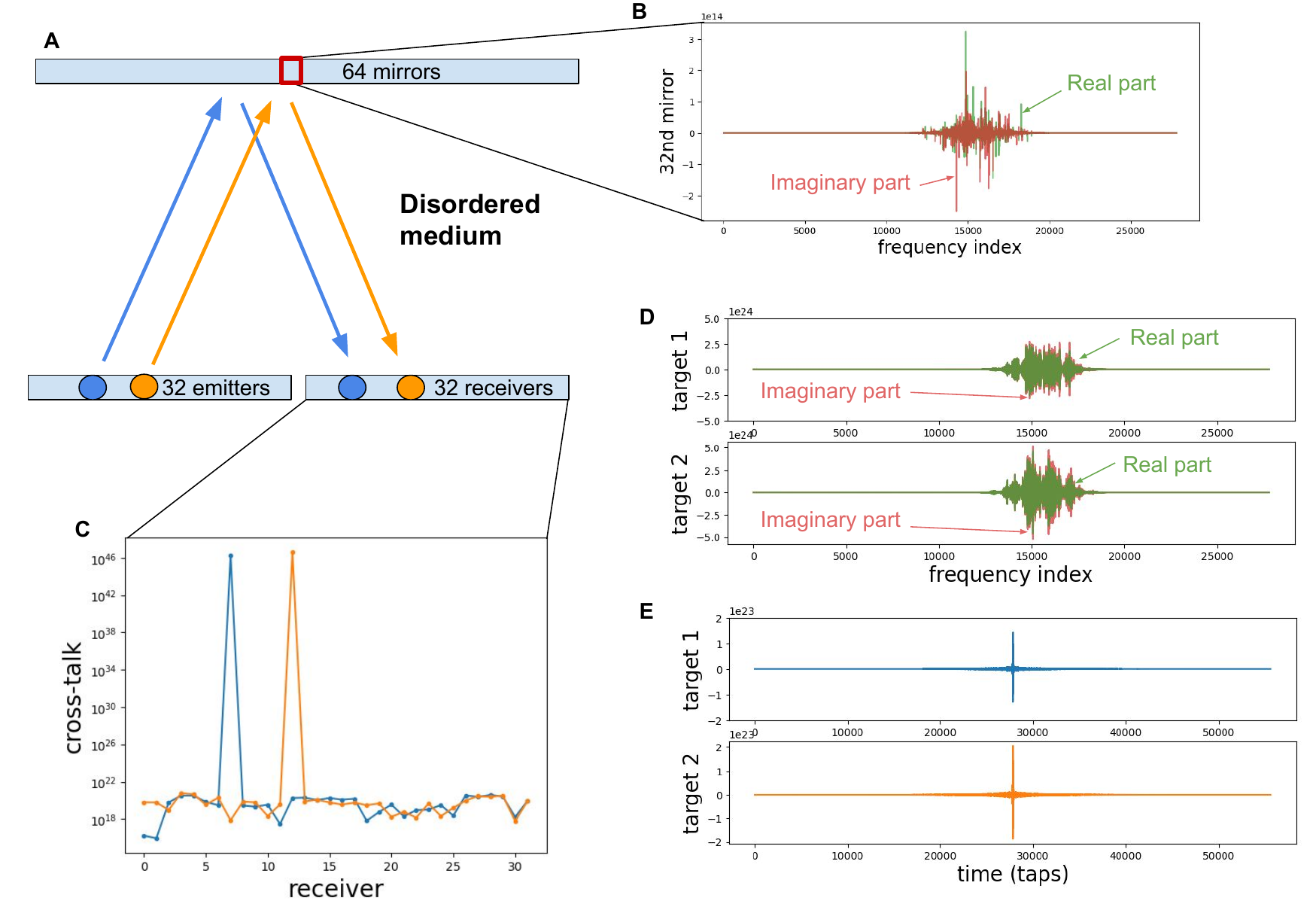}
\caption{Acoustic design for the cocktail party. a) schematic of the setup, b) example optimal frequency response, c) cross-talk, d) frequency response at targets, e) time signal at target.} \label{cap}
\end{figure}

\section{Problem setup}
Similar to~\cite{Bourdeloux24}, we consider the MU-MIMO system shown in Fig.~1(a), where two emitters send a signal to their corresponding receivers. We consider the experimental setup from~\cite{Montaldo04}, where the Green's functions from 32 emitters to 64 mirrors, and from 64 mirrors to 32 receivers, are experimentally measured in disordered media of a forest of rods. The vertical length of the setup is much larger than the physical characteristic length of the setup. Therefore, a two dimensional model suffices.

\section{Time reversal is optimal for focusing at a given time and location}

The intuitive time reversal design is optimal for SISO, where the signal is focused at a given time and a given location~\cite{delhougne16}. Using notations of~\cite{Bourdeloux24}, the contribution to focusing at time $T$ of the path from emitter $E$ to receiver $R$ going through $M_j$ is the temporal convolution
\begin{equation}\label{eq:conv}
    s^j_R(T)= \left[G_{E\rightarrow M_j} \circledast G_{M_j\rightarrow R} \circledast r_{M_j}\right](T) = \int dt \left(G_{E\rightarrow M_j} \circledast G_{M_j\rightarrow R}\right)(T-t)r_{M_j}(t),
\end{equation}
where $G$ are the appropriate Green's functions, $\circledast$ denotes the associative time convolution, and $r_{M_j}$ is the convolutional filter of the $j$-th metasurface element (or mirror). Recognizing an inner product in the right-hand term of Eq.~\ref{eq:conv}, by Chauchy-Schwarz inequality, the contribution is maximum when the two terms are colinear
\begin{equation}\label{eq:tr}
    r^{TR}_{M_j}(t)\propto\left(G_{E\rightarrow M_j} \circledast G_{M_j\rightarrow R}\right)(T-t) , \forall t\in[0,T].
\end{equation}
We recognize the time reversal design in the right-hand term of Eq.~\ref{eq:tr}, which proves the optimality of time reversal design for SISO. Assuming that the signal at the target is dominated by the paths going through the reconfigurable metasurface, and noting $\hat{G}_{E_i M_j R_i}$ the Fourier transform of $\left(G_{E\rightarrow M_j} \circledast G_{M_j\rightarrow R}\right)$, the optimal frequency domain signal at the target using the time reversal from $E_i$ to $R_i$ is
\begin{equation}
    \hat{s}^{TR}_{R_i}(\nu)=\sum_j  \hat{G}_{E_i M_j R_i}(\nu) \hat{r}^{TR}_{M_j}(\nu), \forall \nu,
\end{equation} where $\hat{r}^{TR}_{M_j}$ is the frequency response of the $j$-th mirror from Eq.~\ref{eq:tr}, and the sum is over all the metasurface mirrors. More generally, we define $\hat{G}_{E_i \cdot R_i}$ the vector $(\hat{G}_{E_i M_j R_i})$ , and $\hat{r}_{\cdot}$ the vector $(\hat{r}_{M_j})$, both for all $j$ at a given frequency $\nu$ so that $\hat{s}_{R_i}=(\hat{G}_{E_i \cdot R_i})^T \hat{r}_{\cdot}$.

\section{Least square design formulation in frequency domain}

At a given frequency, for each emitter-receiver $(E_i, R_i)$, we want the signal from $E_i$ to be 
\begin{equation}
    (\hat{G}_{E_i \cdot R_i})^T \hat{r}_{\cdot}=\hat{s}^{TR}_{R_i}
\end{equation} at the target receiver $R_i$ and
\begin{equation}
    (\hat{G}_{E_i \cdot R_{-i}})^T \hat{r}_{\cdot}=0
\end{equation}
at the other receivers (denoted by the index $-i$). This defines the objective frequency response at the receivers $b_i$, and the matrix $A_i$ stacking $(\hat{G}_{E_i \cdot R_i})^T$ vertically for each receivers. For multiple pairs $(E_i, R_i), i=1\cdots n$ we define $A_i$ and $b_i$ for $i=1\cdots n$ and stack them vertically to obtain the objective equality
\begin{equation}
    A\hat{r}_{\cdot} = b
\end{equation} where A is usually a tall matrix. We solve this system in a least square sense with a Ridge regression using its closed form solution for each frequency with the normal equation $(A^TA+\alpha I)\hat{r}_{\cdot}=A^Tb$. $\alpha$ is the coefficient of the Ridge regularization, which may help to avoid overfitting the noise in the data. Note that this framework seamlessly extends to an arbitrary number of pairs in MU-MIMO, and arbitrary $n$ emitters to $n$ receivers communication using the superposition principle. Therefore, we can find the optimal frequency response of the reconfigurable metasurface in a scalable way in space and independently (and potentially in parallel) for each frequency.

\section{Results}
Using the closed form solution of this least square problem, we optimized an acoustic reconfigurable metasurface so that emitter 20 talks to receiver 13 and emitter 24 talks to receiver 17.
Fig.~1(b) shows an the example designed frequency response of the 32nd metasurface element. Note the presence of sharp variations in function of the frequency which may come from overfitting noise in the measurement; here, the fits were performed without regularization ($\alpha=0$). Fig.~1(c) shows the cross-talk, the integral over time of the intensity of the signal received by each receivers in blue from emitter 20 and orange for emitter 24. We see that the cross-talk from this optimal frequency response is almost zero to machine precision. This very strong performance should leave the reader skeptical of a potential overfit of the noise in the Green's functions' data. We leave the analysis of the design robustness to noise to future work. In Fig.~1(d and e), the frequency and time responses, respectively, show a clean pulse for both receivers.

\section{Conclusion}

We introduced a least-square formulation of the MU-MIMO design problem that we solved in closed form in the frequency domain using all the spatiotemporal degrees of freedom of an acoustic reconfigurable metasurface. In the frequency domain, solving for each frequency independently makes the design method efficient and scalable to large systems. In future work, we will design the  corresponding metasurface convolutional filters using algorithms such as the Parks-MacClellan algorithm~\cite{McClellan75} for an experimental validation, and analyze potential overfit.


{\small

\begin{thebibliography}{10}
\setlength{\itemsep}{-1ex}


\bibitem{Bourdeloux24}
C. Bourdeloux, M. Fink, and F. Lemoult, ``A solution to the Cocktail Party Problem: a time reversal active metasurface for multipoint focusing,'' {\itshape Physical Review A,} (in press), 2024.

\bibitem{Arons92}
B. Arons ``A review of the cocktail party effect,'' J. Am. Voice IO Soc. 12, 35–50 (1992).

\bibitem{Assouar18}
B. Assouar et al. ``Acoustic metasurfaces,'' Nat. Rev. Mater. 3, 460–472 (2018).

\bibitem{Sheng06}
P. Sheng, Introduction to wave scattering, localization, and mesoscopic phenomena. (Springer, 2006).

\bibitem{Kaina14}
N. Kaina, et al. ``Shaping complex microwave fields in reverberating media with binary tunable metasurfaces,'' Scientific reports 4.1 (2014): 6693.

\bibitem{Montaldo04}
Montaldo et al. ``Real time inverse filter focusing through iterative time reversal,'' The Journal of the Acoustical Society of America 115.2 (2004): 768-775. 


\bibitem{Zhang24}
H. Zhang et al. ``Optimizing multi-user indoor sound communications with acoustic reconfigurable metasurfaces,'' Nature Communications 15.1 (2024): 1270.

\bibitem{McClellan75}
L. Rabiner, J. McClellan, and T. Parks.``FIR digital filter design techniques using weighted Chebyshev approximation,'' Proceedings of the IEEE 63.4 (1975): 595-610.

\bibitem{delhougne16}
P. Del Hougne, et al. "Spatiotemporal wave front shaping in a microwave cavity." Physical review letters 117.13 (2016): 134302. 

\end{thebibliography}

}
\end{document}